\documentclass[twocolumn,aps,prc,superscriptaddress,showpacs,floatfix]{revtex4}
\usepackage{amssymb}
\usepackage{amsmath}
\usepackage{graphicx}
\usepackage{longtable}
\usepackage[normalem]{ulem}
\usepackage[dvips]{color}

\renewcommand\sout{\bgroup \color{red} \ULdepth=-.5ex \ULset}

\begin{document}

\title{Higher order bulk characteristic parameters of asymmetric nuclear
matter}
\author{Lie-Wen Chen}
\affiliation{Department of Physics, Shanghai Jiao Tong University, Shanghai 200240, China}
\affiliation{Center of Theoretical Nuclear Physics, National Laboratory of Heavy Ion
Accelerator, Lanzhou 730000, China}
\date{\today }

\begin{abstract}
The bulk parameters characterizing the energy of symmetric nuclear
matter and the symmetry energy defined at normal nuclear density
$\rho_0 $ provide important information on the equation of state
(EOS) of isospin asymmetric nuclear matter. While significant
progress has been made in determining some lower order bulk
characteristic parameters, such as the energy $E_0(\rho_0)$ and
incompressibility $K_0$ of symmetric nuclear matter as well as the
symmetry energy $E_{sym}(\rho_0)$ and its slope parameter $L$, yet
the higher order bulk characteristic parameters are still poorly
known. Here, we analyze the correlations between the lower and
higher order bulk characteristic parameters within the framework of
Skyrme Hartree-Fock energy density functional and then estimate the
values of some higher order bulk characteristic parameters. In
particular, we obtain $J_0=-355 \pm 95$ MeV and $I_0=1473 \pm 680$
MeV for the third-order and fourth-order derivative parameters of
symmetric nuclear matter at $\rho_0 $ and $K_{sym} = -100 \pm 165$
MeV, $J_{sym} = 224 \pm 385$ MeV, $I_{sym} = -1309 \pm 2025$ MeV for
the curvature parameter, third-order and fourth-order derivative
parameters of the symmetry energy at $\rho_0 $, using the empirical
constraints on $E_0(\rho_0)$, $K_0$, $E_{sym}(\rho_0)$, $L$, and the
isoscalar and isovector nucleon effective masses. Furthermore, our
results indicate that the three parameters $E_0(\rho_0)$, $K_0$, and
$J_0$ can reasonably characterize the EOS of symmetric nuclear
matter up to $2\rho_0 $ while the symmetry energy up to $2\rho_0 $
can be well described by $E_{sym}(\rho_0)$, $L$, and $K_{sym}$.
\end{abstract}

\maketitle

\section{Introduction}

The equation of state (EOS) of isospin asymmetric nuclear matter,
especially its isospin dependent part which is essentially
characterized by the nuclear symmetry energy, is important for
understanding not only the structure of radioactive nuclei, the
reaction dynamics induced by rare isotopes, and the liquid-gas phase
transition in asymmetric nuclear matter, but also many
critical issues in astrophysics \cite%
{LiBA98,Dan02a,Lat04,Ste05a,Bar05,CKLY07,LCK08}. The nuclear matter
EOS is conventionally defined as the binding energy per nucleon as a
function of the density and a number of bulk parameters defined at
normal nuclear density $\rho_0 $ are usually introduced to
characterize the energy of symmetric nuclear matter and the nuclear
symmetry energy. For example, the energy $E_0(\rho_0)$ and
incompressibility $K_0$ of symmetric nuclear matter are the two
lowest order bulk parameters for the EOS of symmetric nuclear matter
while the symmetry energy $E_{sym}(\rho_0)$ and its slope parameter
$L $ are the two lowest order bulk parameters of the nuclear
symmetry energy. The bulk parameters defined at $\rho_0 $ provide
important information on sub- and supra-saturation density behaviors
of the EOS of isospin asymmetric nuclear matter.

While significant progress has been made in determining some lower
order bulk characteristic parameters of asymmetric nuclear matter,
such as $E_0(\rho_0)$, $K_0$, $E_{sym}(\rho_0)$ and $L$~\cite%
{Bar05,CKLY07,LCK08,You99,Che10,Xu10}, yet the higher order bulk
characteristic parameters are still poorly known. Actually, there is
so far even not any direct experimental information on the
third-order derivative parameter $J_0$ of symmetric nuclear matter
at $\rho_0 $ and the symmetry energy curvature parameter $K_{sym}$.
However, the higher order bulk characteristic parameters have been
shown to be closely related to some important issues in nuclear
physics and astrophysics, such as the determination of the isobaric
incompressibility of asymmetric nuclear matter~\cite{Che09a,Che09b}
and the core-crust transition density and pressure in neutron
stars~\cite{Xu09a,Xu09b,Duc10}.

Theoretically, if the form of an energy density functional and its
parameters are given, then the EOS of isospin asymmetric nuclear
matter can be calculated and thus the bulk characteristic parameters
at any orders can be easily obtained. Usually, the empirical values
of some lower order bulk characteristic parameters, such as
$E_0(\rho_0)$, $K_0$, $E_{sym}(\rho_0)$ and $L$, are used to
constrain the parameters of an energy density functional. In such a
way, while different energy density functionals usually predict
similar results for the lower order bulk characteristic parameters,
they could give very different predictions for the higher order bulk
characteristic parameters. Therefore, the higher order bulk
characteristic parameters can be sensitive to the energy density
functional form and useful for constraining the energy density
functional and its parameters.

In the present work, we analyze the correlations between the lower
and higher order bulk characteristic parameters of asymmetric
nuclear matter within the framework of Skyrme Hartree-Fock energy
density functional. Using the empirical constraints on the lower
order bulk characteristic parameters and other macroscopic
properties of asymmetric nuclear matter, we then estimate the values
of some higher order bulk characteristic parameters.

The paper is organized as follows. We discuss the general properties
of asymmetric nuclear matter in Section \ref{EOS}, and then
introduce the Skyrme Hartree-Fock energy density functional in
Section \ref{SHF}. The results and discussions are presented in
Section \ref{result}. A summary is then given in Section
\ref{summary}.

\section{Equation of state of asymmetric nuclear matter}

\label{EOS}

The EOS of isospin asymmetric nuclear matter, given by its binding energy
per nucleon, can be expanded to $2$nd-order in isospin asymmetry $\delta $
as
\begin{equation}
E(\rho ,\delta )=E_{0}(\rho )+E_{\mathrm{sym}}(\rho )\delta ^{2}+O(\delta
^{4}),  \label{EOSANM}
\end{equation}%
where $\rho =\rho _{n}+\rho _{p}$ is the baryon density with $\rho _{n}$ and
$\rho _{p}$ denoting the neutron and proton densities, respectively; $\delta
=(\rho _{n}-\rho _{p})/(\rho _{p}+\rho _{n})$ is the isospin asymmetry; $%
E_{0}(\rho )=E(\rho ,\delta =0)$ is the binding energy per nucleon in
symmetric nuclear matter, and the nuclear symmetry energy is expressed as
\begin{equation}
E_{\mathrm{sym}}(\rho )=\frac{1}{2!}\frac{\partial ^{2}E(\rho ,\delta )}{%
\partial \delta ^{2}}|_{\delta =0}.  \label{Esym}
\end{equation}%
The absence of odd-order terms in $\delta $ in Eq. (\ref{EOSANM}) is due to
the exchange symmetry between protons and neutrons in nuclear matter when
one neglects the Coulomb interaction and assumes the charge symmetry of
nuclear forces. Neglecting the contribution from higher-order terms in Eq. (%
\ref{EOSANM}) leads to the well-known empirical parabolic law for the EOS of
asymmetric nuclear matter, which has been verified by all many-body theories
to date, at least for densities up to moderate values \cite{LCK08}. As a
good approximation, the density-dependent symmetry energy $E_{\mathrm{sym}%
}(\rho )$ can thus be extracted from the parabolic approximation of $E_{%
\mathrm{sym}}(\rho )\approx E(\rho ,\delta =1)-E(\rho ,\delta =0)$.

Around the nuclear matter saturation density $\rho _{0}$, the binding energy
per nucleon in symmetric nuclear matter $E_{0}(\rho )$\ can be expanded,
e.g., up to $4$th-order in density, as
\begin{equation}
E_{0}(\rho )=E_{0}(\rho _{0})+\frac{K_{0}}{2!}\chi ^{2}+\frac{J_{0}}{3!}\chi
^{3}+\frac{I_{0}}{4!}\chi ^{4}+O(\chi ^{5}),  \label{E0}
\end{equation}%
where $\chi $ is a dimensionless variable characterizing the deviations of
the density from the saturation density $\rho _{0}$ of symmetric nuclear
matter and it is conventionally defined as%
\begin{equation}
\chi =\frac{\rho -\rho _{0}}{3\rho _{0}}.  \label{chi}
\end{equation}%
The first term $E_{0}(\rho _{0})$ on the right-hand-side (r.h.s) of Eq. (\ref%
{E0}) is the binding energy per nucleon in symmetric nuclear matter at the
saturation density $\rho _{0}$ and the coefficients of other terms are
\begin{eqnarray}
K_{0} &=&9\rho _{0}^{2}\frac{d^{2}E_{0}(\rho )}{d\rho ^{2}}|_{\rho =\rho
_{0}}, \\
J_{0} &=&27\rho _{0}^{3}\frac{d^{3}E_{0}(\rho )}{d\rho ^{3}}|_{\rho =\rho
_{0}}, \\
I_{0} &=&81\rho _{0}^{4}\frac{d^{4}E_{0}(\rho )}{d\rho ^{4}}|_{\rho =\rho
_{0}}.
\end{eqnarray}%
The coefficient $K_{0}$ is the incompressibility coefficient of
symmetric nuclear matter and it characterizes the curvature of
$E_{0}(\rho )$ at $\rho _{0}$. The coefficients $J_{0}$ and $I_{0}$
correspond to the $3$rd-order and $4$th-order derivative parameters
of symmetric nuclear matter, respectively.

Around the normal nuclear density $\rho _{0}$, the nuclear symmetry energy $%
E_{\mathrm{sym}}(\rho )$\ can be similarly expanded, e.g., up to $4$th-order
in $\chi $, as
\begin{eqnarray}
E_{\mathrm{sym}}(\rho ) &=&E_{\mathrm{sym}}(\rho _{0})+L\chi +\frac{K_{%
\mathrm{sym}}}{2!}\chi ^{2}  \notag \\
&&+\frac{J_{\mathrm{sym}}}{3!}\chi ^{3}+\frac{I_{\mathrm{sym}}}{4!}\chi
^{4}+O(\chi ^{5}),  \label{EsymExpand}
\end{eqnarray}%
where $L$, $K_{\mathrm{sym}}$, $J_{\mathrm{sym}}$ and
$I_{\mathrm{sym}}$ are the slope parameter, curvature
parameter,curvature parameter, $3$rd-order and $4$th-order
derivative parameters of the nuclear symmetry energy at $\rho _{0}$,
i.e.,
\begin{eqnarray}
L &=&3\rho _{0}\frac{dE_{\mathrm{sym}}(\rho )}{\partial \rho }|_{\rho =\rho
_{0}},  \label{L} \\
K_{\mathrm{sym}} &=&9\rho _{0}^{2}\frac{d^{2}E_{\mathrm{sym}}(\rho )}{%
\partial \rho ^{2}}|_{\rho =\rho _{0}},  \label{Ksym} \\
J_{\mathrm{sym}} &=&27\rho _{0}^{3}\frac{d^{3}E_{\mathrm{sym}}(\rho )}{%
\partial \rho ^{3}}|_{\rho =\rho _{0}}, \\
I_{\mathrm{sym}} &=&81\rho _{0}^{4}\frac{d^{4}E_{\mathrm{sym}}(\rho )}{%
\partial \rho ^{4}}|_{\rho =\rho _{0}}.
\end{eqnarray}%
The coefficients $L$, $K_{\mathrm{sym}}$, $J_{\mathrm{sym}}$ and $I_{\mathrm{%
sym}}$ characterize the density dependence of the nuclear symmetry energy
around the normal nuclear density $\rho _{0}$, and thus carry important
information on the properties of nuclear symmetry energy at both high and
low densities.

In the above Taylor expansions, we have kept all terms up to $4$th-order in $%
\chi $. The $9$ coefficients, namely, $E_{0}(\rho _{0})$, $K_{0}$, $J_{0}$, $%
I_{0}$, $E_{\mathrm{sym}}(\rho _{0})$, $L$, $K_{\mathrm{sym}}$, $J_{\mathrm{%
sym}}$, $I_{\mathrm{sym}}$, are theoretically well-defined, and they
characterize the EOS of an asymmetric nuclear matter and its density
dependence at the normal nuclear density $\rho _{0}$. Among these
parameters, the lower order bulk characteristic parameters $E_{0}(\rho _{0})$%
, $K_{0}$, $E_{\mathrm{sym}}(\rho _{0})$, and $L $ have been extensively
studied in the literature and significant progress has been made over past
few decades. Based on the empirical constraints on the lower order bulk
characteristic parameters and other macroscopic properties of asymmetric
nuclear matter, we investigate in the following to what extend the higher
order parameters $J_{0}$, $I_{0}$, $K_{\mathrm{sym}}$, $J_{\mathrm{sym}}$,
and $I_{\mathrm{sym}}$ can be constrained within the framework of Skyrme
Hartree-Fock energy density functional.

\section{Skyrme-Hartree-Fock approach and macroscopic properties of
asymmetric nuclear matter}

\label{SHF}

In the standard Skyrme Hartree-Fock model, the nuclear effective interaction
is taken to have a zero-range, density- and momentum-dependent form \cite%
{Cha97}, i.e.,
\begin{eqnarray}
V_{12}(\mathbf{R},\mathbf{r}) &=&t_{0}(1+x_{0}P_{\sigma })\delta (\mathbf{r})
\notag \\
&+&\frac{1}{6}t_{3}(1+x_{3}P_{\sigma })\rho ^{\sigma }(\mathbf{R})\delta (%
\mathbf{r})  \notag \\
&+&\frac{1}{2}t_{1}(1+x_{1}P_{\sigma })(K^{^{\prime }2}\delta (\mathbf{r}%
)+\delta (\mathbf{r})K^{2})  \notag \\
&+&t_{2}(1+x_{2}P_{\sigma })\mathbf{K}^{^{\prime }}\cdot \delta (\mathbf{r})%
\mathbf{K}  \notag \\
&\mathbf{+}&iW_{0}(\mathbf{\sigma }_{1}+\mathbf{\sigma }_{2})\cdot \lbrack
\mathbf{K}^{^{\prime }}\times \delta (\mathbf{r})\mathbf{K]},
\end{eqnarray}%
with $\mathbf{r}=\mathbf{r}_{1}-\mathbf{r}_{2}$ and $\mathbf{R}=(\mathbf{r}%
_{1}+\mathbf{r}_{2})/2$. In the above, the relative momentum operators $%
\mathbf{K}=(\mathbf{\nabla }_{1}-\mathbf{\nabla }_{2})/2i$ and $\mathbf{K}%
^{\prime }=-(\mathbf{\nabla }_{1}-\mathbf{\nabla }_{2})/2i$ act on the wave
function on the right and left, respectively. The quantities $P_{\sigma }$
and $\sigma _{i}$ denote, respectively, the spin exchange operator and Pauli
spin matrices. The $\sigma $, $t_{0}-t_{3}$, $x_{0}-x_{3}$ are the $9$
Skyrme interaction parameters and $W_{0}$ is the spin-orbit coupling
constant. Within the standard form, the EOS of symmetric nuclear matter can
be written as

\begin{eqnarray}
E_{0}(\rho ) &=&\frac{3\hbar ^{2}}{10m}\left( \frac{3\pi ^{2}}{2}\right)
^{2/3}\rho ^{\frac{2}{3}}+\frac{3}{8}t_{0}\rho  \notag \\
&+&\frac{3}{80}\Theta _{s}\left( \frac{3\pi ^{2}}{2}\right) ^{2/3}\rho ^{%
\frac{5}{3}}+\frac{1}{16}t_{3}\rho ^{\sigma +1},
\end{eqnarray}%
with $\Theta _{s}=3t_{1}+(5+4x_{2})t_{2}$. Furthermore, the symmetry energy
can be obtained as

\begin{eqnarray}
E_{\text{\textrm{sym}}}(\rho ) &=&\frac{1}{2}\left( \frac{\partial ^{2}E}{%
\partial \delta ^{2}}\right) _{\delta =0}  \notag \\
&=&\frac{\hbar ^{2}}{6m}\left( \frac{3\pi ^{2}}{2}\right) ^{2/3}\rho ^{\frac{%
2}{3}}-\frac{1}{8}t_{0}(2x_{0}+1)\rho  \notag \\
&-&\frac{1}{24}\left( \frac{3\pi ^{2}}{2}\right) ^{2/3} \Theta _{\mathrm{sym}%
}\rho ^{\frac{5}{3}}  \notag \\
&-&\frac{1}{48}t_{3}(2x_{3}+1)\rho ^{\sigma +1},
\end{eqnarray}%
with $\Theta _{\mathrm{sym}}=3t_{1}x_{1}-t_{2}(4+5x_{2})$.

As shown in Ref.~\cite{Che10}, the $9$ Skyrme interaction parameters, i.e., $%
\sigma $, $t_{0}-t_{3}$, $x_{0}-x_{3}$ can be expressed analytically
in terms of $9$ macroscopic quantities $\rho _{0}$, $E_{0}(\rho
_{0})$, the incompressibility $K_{0}$, the isoscalar effective mass
$m_{s,0}^{\ast }$,
the isovector effective mass $m_{v,0}^{\ast }$, $E_{\text{\textrm{sym}}}({%
\rho _{0}})$, $L$, gradient coefficient $G_{S}$, and symmetry-gradient
coefficient $G_{V}$, i.e.,
\begin{eqnarray}
t{_{0}} &=&4\alpha /(3{\rho _{0}}) \\
x{_{0}} &=&3(y-1)E_{\text{\textrm{sym}}}^{\mathrm{loc}}({\rho _{0}})/\alpha
-1/2 \\
t{_{3}} &=&16\beta /\left[ {\rho _{0}}^{\gamma }(\gamma +1)\right] \\
x{_{3}} &=&-3y(\gamma +1)E_{\text{\textrm{sym}}}^{\mathrm{loc}}({\rho _{0}}%
)/(2\beta )-1/2 \\
t_{1} &=&20C/\left[ 9{\rho _{0}(}k_{\mathrm{F}}^{0})^{2}\right] +8G_{S}/3 \\
t_{2} &=&\frac{4(25C-18D)}{9{\rho _{0}(}k_{\mathrm{F}}^{0})^{2}}-\frac{%
8(G_{S}+2G_{V})}{3} \\
x_{1} &=&\left[ 12G_{V}-4G_{S}-\frac{6D}{{\rho _{0}(}k_{\mathrm{F}}^{0})^{2}}%
\right] /(3t_{1}) \\
x_{2} &=&\left[ 20G_{V}+4G_{S}-\frac{5(16C-18D)}{3{\rho _{0}(}k_{\mathrm{F}%
}^{0})^{2}}\right] /(3t_{2}) \\
\text{\ }\sigma &=&\gamma -1
\end{eqnarray}%
where the parameters $C$, $D$, $\alpha $, $\beta $, $\gamma $, $E_{\text{%
\textrm{sym}}}^{\mathrm{loc}}({\rho _{0}})$, and $y$ are defined as

\begin{eqnarray}
C &=&\frac{m-m_{s,0}^{\ast }}{m_{s,0}^{\ast }}E_{\mathrm{kin}}^{0} \\
D &=&\frac{5}{9}E_{\mathrm{kin}}^{0}\left( 4\frac{m}{m_{s,0}^{\ast }}-3\frac{%
m}{m_{v,0}^{\ast }}-1\right) \\
\alpha &=&-\frac{4}{3}E_{\mathrm{kin}}^{0}-\frac{10}{3}C-\frac{2}{3}(E_{%
\mathrm{kin}}^{0}-3E_{0}(\rho _{0})-2C)  \notag \\
&&\times \frac{K_{0}+2E_{\mathrm{kin}}^{0}-10C}{K_{0}+9E_{0}(\rho _{0})-E_{%
\mathrm{kin}}^{0}-4C} \\
\beta &=&(\frac{E_{\mathrm{kin}}^{0}}{3}-E_{0}(\rho _{0})-\frac{2}{3}C)
\notag \\
&&\times \frac{K_{0}-9E_{0}(\rho _{0})+5E_{\mathrm{kin}}^{0}-16C}{%
K_{0}+9E_{0}(\rho _{0})-E_{\mathrm{kin}}^{0}-4C} \\
\gamma &=&\frac{K_{0}+2E_{\mathrm{kin}}^{0}-10C}{3E_{\mathrm{kin}%
}^{0}-9E_{0}(\rho _{0})-6C}. \\
E_{\text{\textrm{sym}}}^{\mathrm{loc}}({\rho _{0}}) &=&E_{\text{\textrm{sym}}%
}({\rho _{0}})-E_{\text{\textrm{sym}}}^{\mathrm{kin}}({\rho _{0}})-D \\
y &=&\frac{L-3E_{\text{\textrm{sym}}}({\rho _{0}})+E_{\text{\textrm{\ sym}}%
}^{\mathrm{kin}}({\rho _{0}})-2D}{3(\gamma -1)E_{\text{ \textrm{sym}}}^{%
\mathrm{loc}}({\rho _{0}})}
\end{eqnarray}%
with $k_{\mathrm{F}}^{0}=\left( 1.5\pi ^{2}{\rho _{0}}\right) ^{1/3}$, $E_{%
\mathrm{kin}}^{0}=\frac{3\hbar ^{2}}{10m}\left( \frac{3\pi ^{2}}{2}\right)
^{2/3}\rho _{0}^{2/3}$, and $E_{\text{\textrm{sym}}}^{kin}({\rho _{0}})=%
\frac{\hbar ^{2}}{6m}\left( \frac{3\pi ^{2}}{2}{\rho _{0}}\right) ^{2/3}$.
Furthermore, $G_{S}$ and $G_{V}$ are respectively the gradient and
symmetry-gradient coefficients in the interaction part of the binding
energies for finite nuclei defined as
\begin{equation}
E_{\mathrm{grad}}=G_{S}(\nabla \rho )^{2}/(2{\rho )}-G_{V}\left[ \nabla
(\rho _{n}-\rho _{p})\right] ^{2}/(2{\rho )}.
\end{equation}

\section{Results and discussions}

\label{result}

Based on the formulism in previous sections, one can now estimate
the values of higher order bulk characteristic parameters, such as
$J_0$ and $K_{sym}$
by analyzing their correlations with $\rho _{0}$, $E_{0}(\rho _{0})$, $K_{0}$%
, $m_{s,0}^{\ast }$, $m_{v,0}^{\ast }$, $E_{\text{\textrm{sym}}}({\rho _{0}}%
) $, $L$, $G_{S}$, and $G_{V}$ within the framework of Skyrme Hartree-Fock
energy density functional. As a reference for the correlation analyses
below, we use the MSL0 parameter set~\cite{Che10}, which is obtained by
using the following empirical values for the macroscopic quantities: $\rho
_{0}=0.16$ fm$^{-3}$, $E_{0}(\rho _{0})=-16$ MeV, $K_{0}=230$ MeV, $%
m_{s,0}^{\ast }=0.8m $, $m_{v,0}^{\ast }=0.7m$, $E_{\text{\textrm{sym}}}({%
\rho _{0}})=30$ MeV, and $L=60$ MeV, $G_{V}=5$ MeV$\cdot $fm$^{5}$, and $%
G_{S}=132$ MeV$\cdot $fm$^{5}$. And $W_{0}=133.3$ MeV $\cdot
$fm$^{5}$ is used to fit the neutron $p_{1/2}-p_{3/2}$ splitting in
$^{16}$O. It has been shown~\cite{Che10} that the MSL0 interaction
can describe reasonably (the relative deviation from the data is
less than $2\%$) the binding energies and charge rms radii for a
number of closed-shell or semi-closed-shell nuclei. It should be
pointed out that the MSL0 is only used here as a reference for the
correlation analyses below. Using other Skyrme interactions obtained
from fitting measured binding energies and charge rms radii of
finite nuclei as in usual Skyrme parametrization will not change our
conclusion.

\begin{figure}[tbp]
\includegraphics[scale=0.7]{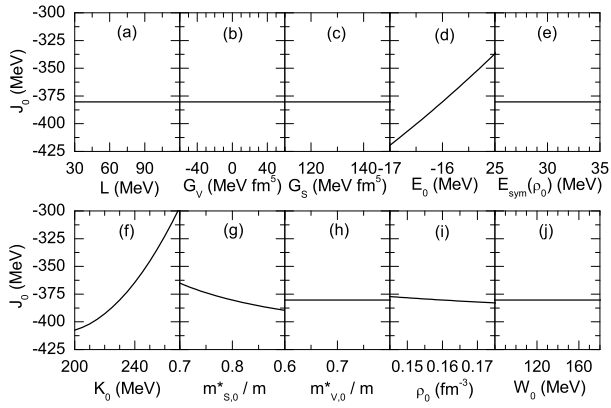}
\caption{$J_{0}$ from SHF with MSL0 by varying individually $L$ (a), $G_{V}$
(b), $G_{S}$ (c), $E_{0}(\protect\rho _{0})$ (d), $E_{\text{\textrm{sym}}}(%
\protect\rho _{0})$ (e), $K_{0}$ (f), $m_{s,0}^{\ast }$ (g), $m_{v,0}^{\ast
} $ (h), $\protect\rho _{0}$ (i), and $W_{0}$ (j).}
\label{XJ0}
\end{figure}

\begin{figure}[tbp]
\includegraphics[scale=0.7]{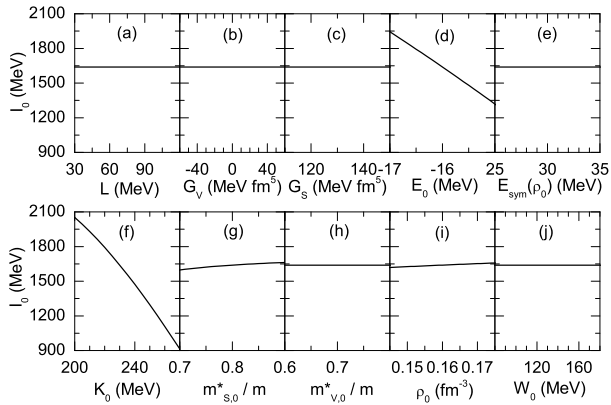}
\caption{Same as Figure \protect\ref{XJ0} but for $I_{0}$.}
\label{XI0}
\end{figure}

To reveal clearly the correlation of higher order bulk characteristic
parameters with each macroscopic quantity, we vary one macroscopic quantity
at a time while keeping all others at their default values in MSL0. Shown in
Fig. \ref{XJ0} is the value of the higher order bulk characteristic
parameter $J_0$. Within the uncertain ranges considered here, the parameter $%
J_0$ exhibits a very strong correlation with $K_{0}$ and $E_{0}(\rho _{0})$.
However, it depends only moderately on $m_{s,0}^{\ast }$ and very weakly on $%
\rho _{0}$ while it displays no dependence on the other parameters $E_{\text{%
\textrm{sym}}}({\rho _{0}})$, $L$, $G_{S}$, $G_{V}$, $m_{v,0}^{\ast }$, and $%
W_{0}$. From the empirical values of $E_0=-16 \pm 1$ MeV, $K_0=240 \pm 20$
MeV, and $m_{s,0}^{\ast }=(0.8 \pm 0.1)m $, we can obtain an estimate of $%
J_0=-355 \pm 95$ MeV. The results of a similar correlation analysis on the
parameter $I_0$ are shown in Fig. \ref{XI0}, and it is seen that the
parameter $I_0$ also exhibits a very strong correlation with $K_{0}$ and $%
E_{0}(\rho _{0})$ and the empirical values of $E_0=-16 \pm 1$ MeV and $%
K_0=240 \pm 20$ MeV lead to an extraction of $I_0=1473 \pm 680$ MeV.
Therefore, the higher order bulk characteristic parameters $J_0$ and
$I_0$ can be determined within relative uncertainties of about
$27\%$ and $46\%$, respectively, in the present analysis.

\begin{figure}[tbp]
\includegraphics[scale=0.7]{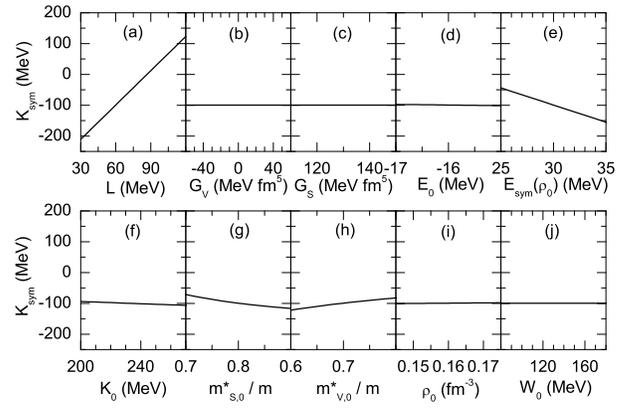}
\caption{Same as Figure \protect\ref{XJ0} but for $K_{sym}$.}
\label{XKsym}
\end{figure}

\begin{figure}[tbp]
\includegraphics[scale=0.7]{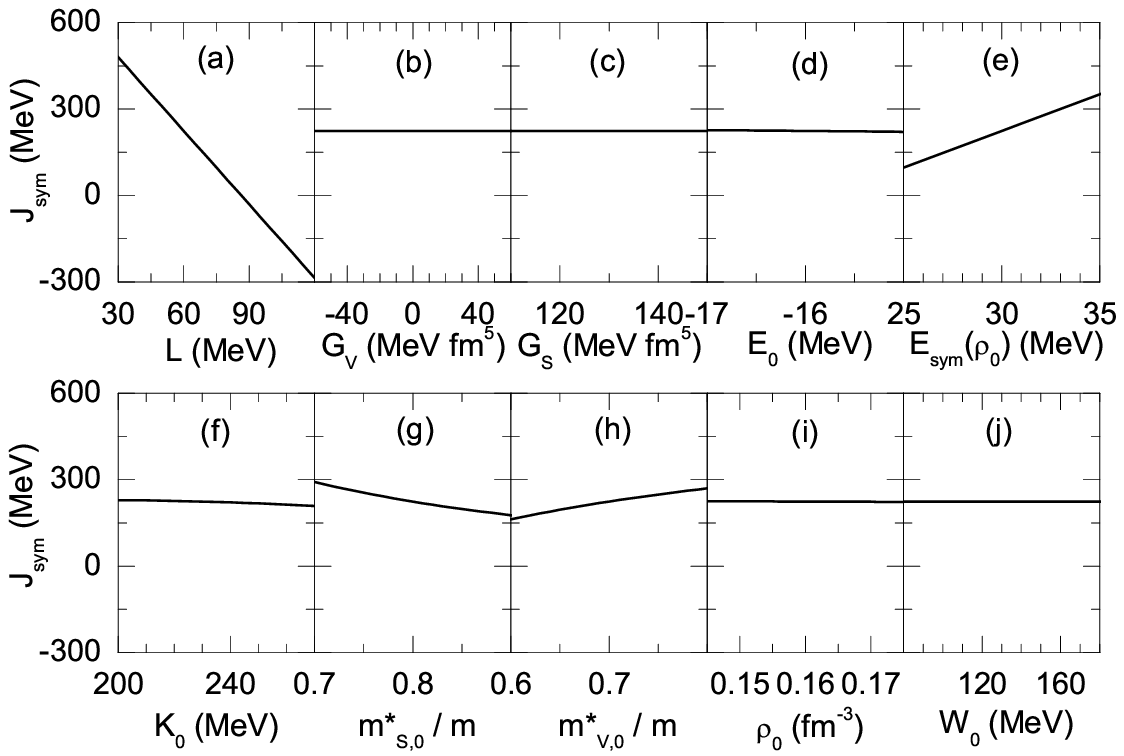}
\caption{Same as Figure \protect\ref{XJ0} but for $J_{sym}$.}
\label{XJsym}
\end{figure}

\begin{figure}[tbp]
\includegraphics[scale=0.7]{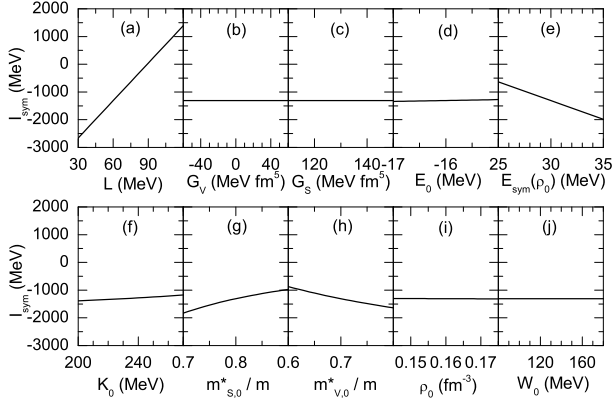}
\caption{Same as Figure \protect\ref{XJ0} but for $I_{sym}$.}
\label{XIsym}
\end{figure}

In Fig. \ref{XKsym} we make a similar correlation analysis as in Fig. \ref%
{XJ0} for the symmetry energy curvature parameter $K_{sym}$. It is clearly
seen that the parameter $K_{sym}$ exhibits a very strong correlation with $L$
and $E_{\text{\textrm{sym}}}({\rho _{0}})$. However, it depends only
moderately on $m_{s,0}^{\ast }$ and $m_{v,0}^{\ast }$ and very weakly on $%
E_{0}(\rho _{0})$, $K_{0}$, and $\rho _{0}$ while it is independent of other
parameters $G_{S}$, $G_{V}$, and $W_{0}$. From the empirical values of $L=60
\pm 30$ MeV and $E_{\text{\textrm{sym}}}(\rho _{0})=30 \pm 5 $MeV, $%
m_{s,0}^{\ast }=(0.8 \pm 0.1)m $, and $m_{s,0}^{\ast }-m_{v,0}^{\ast
}=(0.126 \pm 0.051)m $~\cite{Xu10}, we can obtain $K_{sym}=-100 \pm 165$
MeV. Furthermore, Figs. \ref{XJsym} and \ref{XIsym} display the results from
similar analysis as in Fig. \ref{XJ0} for the higher order bulk
characteristic parameters $J_{sym}$ and $I_{sym}$, respectively. Similarly
as in the case of $K_{sym}$, they exhibit very strong correlation with $L$
and $E_{\text{\textrm{sym}}}({\rho _{0}})$. The empirical values of $L=60
\pm 30$ MeV, $E_{\text{\textrm{sym}}}(\rho _{0})=30 \pm 5 $MeV, $%
m_{s,0}^{\ast }=(0.8 \pm 0.1)m $, and $m_{s,0}^{\ast }-m_{v,0}^{\ast
}=(0.126 \pm 0.051)m $ lead to the estimates of $J_{sym}=224 \pm
385$ MeV and $I_{sym}=-1309 \pm 2025$ MeV. These results indicate
that the higher order bulk characteristic parameters of the symmetry
energy, i.e., $K_{sym}$, $J_{sym}$ and $I_{sym}$, are largely
uncertain based on our present knowledge in the standard SHF energy
density functional.

\begin{figure}[tbp]
\includegraphics[scale=0.75]{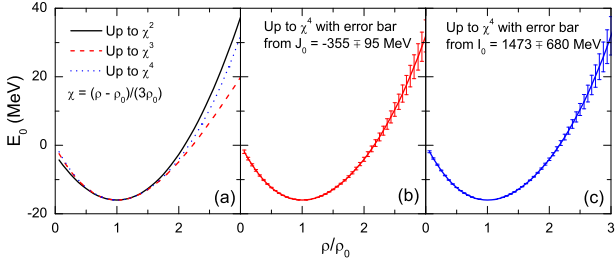}
\caption{(a) Energy per nucleon as a function of density for symmetric
nuclear matter obtained by using Eq. (\protect\ref{E0}) including terms up
to $\protect\chi^{2}$, $\protect\chi^{3}$, and $\protect\chi^{4}$,
respectively. (b) Energy per nucleon as a function of density for symmetric
nuclear matter obtained by using Eq. (\protect\ref{E0}) including terms up
to $\protect\chi^{4}$ with error bars due to the uncertainty of $J_{0}$. (c)
Same as (b) but with error bars due to the uncertainty of $I_{0}$. ($E_{0}(%
\protect\rho _{0})=-16$ MeV and $K_{0}=230$ MeV have been assumed in
the calculations.)} \label{EOS0KJI}
\end{figure}

Using the estimated values of the higher order bulk characteristic
parameters, one can see how they influence the EOS of asymmetric nuclear
matter. Shown in Fig. \ref{EOS0KJI} is the case for symmetric nuclear
matter. Fig. \ref{EOS0KJI}(a) shows the results obtained by using Eq. (\ref%
{E0}) including terms up to $\chi^{2}$, $\chi^{3}$, and $\chi^{4}$,
respectively. One can see that Eq. (\ref{E0}) with terms up to
$\chi^{2}$ [i.e., the parabolic approximation] already gives a
convergent prediction for the EOS of symmetric nuclear matter from
about $0.5\rho_{0}$ to $1.5\rho_{0}$. The higher order terms of
$\chi^{3}$ and $\chi^{4}$ with the characteristic parameters $J_{0}$
and $I_{0}$ in Eq. (\ref{E0}) significantly improves the
approximation to the EOS at low densities and that up to about
$2\rho_{0}$. To describe reasonably the EOS of symmetric nuclear
matter above $2\rho_{0}$, one needs to include higher order terms in
$\chi$. These results are consistent with those obtained in
Ref.~\cite{Che09b} where the MDI interaction has been used. Figs.
\ref{EOS0KJI}(b) and (c) display how the uncertainties of $J_{0}$
and $I_{0}$ affect the EOS of symmetric nuclear matter. It is seen
that their uncertainties have only minor influence on the
subsaturation density behaviors but significantly affect the EOS at
higher densities (above about $2\rho_{0}$). These results imply
that the EOS of symmetric nuclear matter in densities up to about $%
2\rho_{0}$ can already be well described by the three bulk characteristic
parameters $E_{0}(\rho _{0})$, $K_{0}$, and $J_{0}$.

\begin{figure}[tbp]
\includegraphics[scale=1.1]{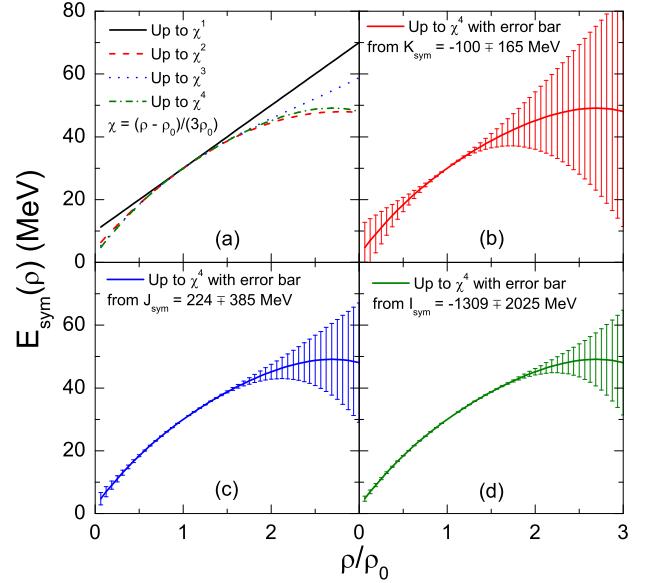}
\caption{(a) Symmetry energy as a function of density obtained by using Eq. (%
\protect\ref{EsymExpand}) including terms up to $\protect\chi$, $\protect\chi%
^{2}$, $\protect\chi^{3}$, and $\protect\chi^{4}$, respectively. (b)
Symmetry energy as a function of density obtained by using Eq. (\protect\ref%
{EsymExpand}) including terms up to $\protect\chi^{4}$ with error
bars due to the uncertainty of $K_{sym}$. (c) Same as (b) but with
error bars due to the uncertainty of $J_{sym}$. (d) Same as (b) but
with error bars due to the uncertainty of $I_{sym}$.(
$E_{\text{\textrm{sym}}}({\protect\rho _{0}})=30$ MeV and $L=60$ MeV
have been assumed in the calculations.)} \label{EsymLKJI}
\end{figure}

Figure 7 displays how the higher order bulk characteristic
parameters $K_{sym}$, $J_{sym}$ and $I_{sym}$ influence the density
dependence of symmetry energy. Figure 7(a) shows the results
obtained by using Eq. (\ref{EsymExpand}) including terms up to
$\chi$, $\chi^{2}$, $\chi^{3}$, and $\chi^{4}$, respectively. It is
seen that Eq. (\ref{EsymExpand}) with terms up to $\chi^{2}$ can
already give reasonably a convergent result for the density
dependence of symmetry energy up to about $2\rho_{0}$. The higher
order terms of $\chi^{3}$ and $\chi^{4}$ with the characteristic
parameters $J_{sym}$ and $I_{sym}$ in Eq. (\ref{EsymExpand})
significantly improve the approximation to the symmetry energy up to
about $3\rho_{0}$. Figures 7(b), 7(c) and 7(d) display how the
uncertainties of $K_{sym}$, $J_{sym}$ and $I_{sym}$ affect the
density dependence of symmetry energy. One can see that the
uncertainty of the parameter $K_{sym}$ significantly affects both
the sub- and supra-saturation density behaviors of the symmetry
energy while the uncertainties of $J_{sym}$ and $I_{sym}$ only have
significant influence on the symmetry energy at higher densities
(above about $2\rho_{0}$). These features indicate that the three
bulk characteristic parameters $E_{\text{\textrm{sym}}}({\rho
_{0}})$, $L$, and $K_{sym}$ essentially determine the symmetry
energy with the density up to about $2\rho_{0}$.





\section{Summary}

\label{summary}

We have analyzed the correlations between the lower and higher order
bulk characteristic parameters of asymmetric nuclear matter within
the framework of Skyrme Hartree-Fock energy density functional.
Based on these correlations, we have estimated the values of some
higher order bulk characteristic parameters. In particular, we have
obtained $J_0=-355 \pm 95$ MeV, $I_0=1473 \pm 680$ MeV, $K_{sym} =
-100 \pm 165$ MeV, $J_{sym} = 224 \pm 385$ MeV, and $I_{sym} = -1309
\pm 2025$ MeV using the empirical constraints on $E_0(\rho_0)$,
$K_0$, $E_{sym}(\rho_0)$, $L$, and the isoscalar and isovector
nucleon effective masses. Our results indicate that the three bulk
characteristic parameters $E_{0}(\rho _{0})$, $K_{0}$, and $J_{0}$
essentially determine the EOS of symmetric nuclear matter in
densities up to about $2\rho_{0}$ while the three bulk
characteristic parameters $E_{\text{\textrm{sym}}}({\rho _{0}})$,
$L$, and $K_{sym}$ well characterize the symmetry energy in
densities up to about $2\rho_{0}$. For higher density (above
$2\rho_{0}$) behaviors of the EOS of asymmetric nuclear matter, the
higher order bulk characteristic parameters become important.

In the present work we have estimated the higher order bulk
characteristic parameters only based on the standard SHF energy
density functional. It will be interesting to see how our results
change if different energy-density functionals are used and these
studies are in progress. The experimental information or model
independent prediction on higher order bulk characteristic
parameters of asymmetric nuclear matter is expected to put important
constraints on the nuclear energy density functional form and its
parameters.

\begin{acknowledgments}
The author thanks Che Ming Ko, Bao-An Li, Chang Xu, and Jun Xu for
helpful discussions. This work was supported in part by the NNSF of
China under Grant No. 10975097, Shanghai Rising-Star Program under
Grant No. 06QA14024, the National Basic Research Program of China
(973 Program) under Contract No. 2007CB815004.
\end{acknowledgments}

\end{document}